\documentclass{ws-procs9x6}

\newcommand{\be}{\begin{equation}}
\newcommand{\ee}{\end{equation}}
\newcommand{\bra}{\langle}
\newcommand{\ket}{\rangle}
\newcommand{\Tr}{{\mathrm{Tr}}}
\newcommand{\half}{\frac{1}{2}}
\newcommand{\pv}{{\mathbf p}}

\newcommand{\vecnul}{{\mathbf 0}}
\newcommand{\om}{\omega}

\begin{document}

\title{
Transport coefficients and the 2PI effective action in the large $N$ limit\footnote{
\uppercase{B}ased on presentations at
\uppercase{S}trong and \uppercase{E}lectroweak \uppercase{M}atter
(\uppercase{SEWM}04), \uppercase{H}elsinki, \uppercase{F}inland,
\uppercase{J}une 16-19 2004, and the \uppercase{W}orkshop on
\uppercase{QCD} in \uppercase{E}xtreme \uppercase{E}nvironments,
\uppercase{HEP} \uppercase{D}ivision, \uppercase{A}rgonne
\uppercase{N}ational \uppercase{L}aboratory, \uppercase{USA},
\uppercase{J}une 29-\uppercase{J}uly 3 2004.
}}

\author{Gert Aarts and Jose M. Martinez Resco}

\address{
Department of Physics, The Ohio State University\\
174 West 18th Avenue, Columbus, OH 43210, USA\\ 
aarts@mps.ohio-state.edu, marej@mps.ohio-state.edu}

\maketitle

\abstracts{
 We discuss the computation of transport coefficients in large $N_{f}$ QCD
and the $O(N)$ model for massive particles.  The calculation is organized
using the $1/N$ expansion of the 2PI effective action to next-to-leading
order. For the gauge theory, we verify gauge fixing independence and
consistency with the Ward identity. In the gauge theory, we find a
nontrivial dependence on the fermion mass.
 }


\section{Motivation}

In understanding the evolution of quantum fields out of equilibrium,
thermalization plays an important role. For long enough times, the
dynamics of a typical system can be characterized by transport
coefficients, as they control the relaxation of a system towards
equilibrium. For example, in heavy ion collisions, the presence of an
appreciable shear viscosity would modify the ideal hydrodynamical
description\cite{Kolb:2003dz} of the aftermath of the
collision.\cite{Teaney:2003pb,Muronga:2004sf}

The nonequilibrium evolution of quantum fields and the subsequent
thermalization have been studied using various approximations: mean-field
dynamics (such as Hartree or leading-order large $N$ approximations),
inhomogeneous mean-field theory, the classical approximation, kinetic
theory, dynamics based on truncations of the 2PI or 2PPI effective action,
etc. It is clear that any scheme aimed at describing the long-time
behavior should, among other things, describe correctly the approach to
equilibrium of a typical system. One requirement for an approximation
scheme is therefore to yield sensible results for transport coefficients.
While it is clear that some approaches (mean-field theory, classical
approximation) do not give the correct long-time quantum behavior, for
others this might be more involved.

One approach to the dynamics of quantum fields far from equilibrium that
has been successful in the past few
years\cite{Berges:2000ur}$^-$\cite{Ikeda:2004in} employs the 2PI effective
action.\cite{Cornwall:1974vz} It is therefore interesting to see whether
the 2PI framework gives a reliable description of transport coefficients,
in the limit where a semi-analytical computation can be carried out, i.e.\
in a weak coupling or a large $N$ expansion. Below we review how transport
coefficients can be computed using the 2PI effective action as an
organizational tool.\cite{Aarts:2003bk} We discuss explicitly two theories
in the large $N$ limit: the shear viscosity in the $O(N)$
model\cite{Aarts:2004sd} and the electrical conductivity and shear
viscosity in large $N_f$ QCD with massive fermions.\cite{Aarts:2004}


\section{2PI effective action}

The 2PI effective action offers an approach to deal with the 
infinite hierarchy of correlation functions in field theory, based on a 
variational principle for one- and two-point functions.
For a bosonic field, with vanishing mean field $\bra\phi\ket=0$, the 
effective action is written as\cite{Cornwall:1974vz}
\be
\Gamma[G] = \frac{i}{2}\Tr \ln G^{-1} + \frac{i}{2}\Tr\, G_0^{-1}(G-G_0)
+\Gamma_2[G],
\ee
 where $G_0^{-1}$ is the free inverse propagator. $\Gamma_2[G]$ contains
all two-particle irreducible diagrams without external legs. For fermionic
fields the $\half$'s are replaced by $-1$'s.  Although the attention is
usually focused on the (one- and) two-point functions, the 2PI effective
action is formulated as a generating functional with external sources,
similar to the 1PI effective action, and therefore higher order $n$-point
functions are accessible as well. In particular, there is a four-point 
vertex function obeying a Bethe-Salpeter equation, which in momentum 
space reads
 \be
\Gamma^{(4)}(P,K) = \Lambda(P,K) + \half \int_R \Lambda(P,R) G^2(R) 
\Gamma^{(4)}(R,K).
\ee
 This equation sums ladder diagrams with a kernel $\Lambda$
obtained from the second derivative of $\Gamma_2[G]$. It plays therefore a
crucial role in the computation of transport
coefficients\cite{Aarts:2003bk} in the case that these are dominated by
ladder diagrams, such as in scalar\cite{Jeon:if,Aarts:2004sd}
and large $N_f$ gauge theories\cite{Aarts:2004} to leading order and in
weakly coupled gauge theories to leading logarithmic
order.\cite{ValleBasagoiti:2002ir}

We now apply this to two theories in the large $N$ limit: the $O(N)$ model
and large $N_f$ QCD. We consider the shear viscosity in the $O(N)$ model
with massive particles both at weak coupling as well as in the large $N$
limit. For a weakly-coupled single-component scalar field the shear
viscosity was computed some time ago by Jeon.\cite{Jeon:if} In large $N_f$
QCD, we present the first diagrammatic calculation of the electrical
conductivity and shear viscosity for massive fermions. For massless
fermions, these transport coefficients were computed before by
Moore\cite{Moore:2001fg} using kinetic theory.

\begin{figure}[ht]
\centerline{
\epsfysize=1.2cm\epsfbox{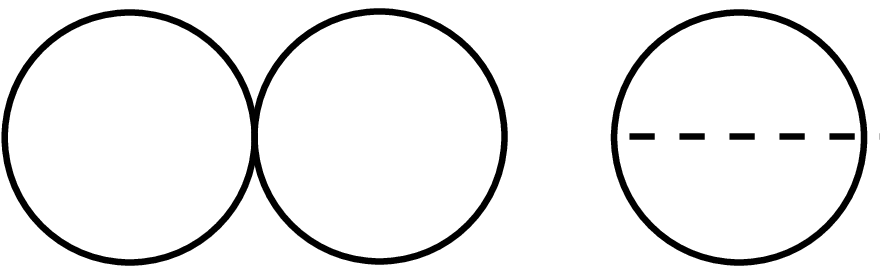}
\hspace*{2cm}
\epsfysize=1.2cm\epsfbox{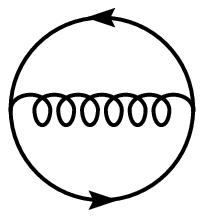}
}   
\caption{2PI contribution to the effective action at LO and NLO in the 
large $N$ limit in the $O(N)$ model (left) and at NLO in large $N_f$ QCD 
(right). The dashed line in the $O(N)$ model sums the chain of bubbles, 
see Fig.\ \ref{figaux}. 
}
\label{figeffaction}
\end{figure}

\begin{figure}[ht]
\centerline{
\epsfysize=0.6cm\epsfbox{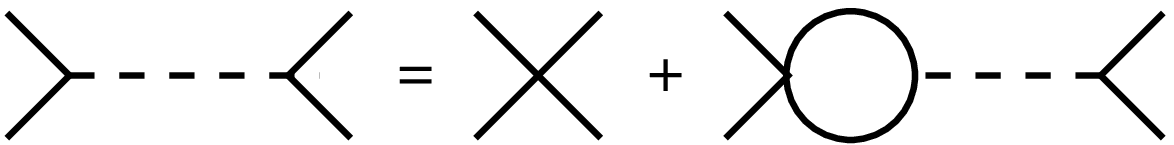}
}   
\caption{
Auxiliary correlator in the $O(N)$ model, summing the chain of bubbles.
}
\label{figaux}
\end{figure}

The starting point is the 2PI contribution to the effective action in the
$1/N$ expansion, shown in Fig.\ \ref{figeffaction}. In the $O(N)$ model
there is a LO and a NLO contribution.\cite{Berges:2001fi,Aarts:2002dj} In
large $N_f$ QCD there is only a contribution at NLO. The chain of bubbles
in the $O(N)$ model is indicated with the dashed line (see Fig.\
\ref{figaux}). In this formulation the scalar and gauge theory are
conveniently similar. We emphasize that all results below are generated
from the graphs in Fig.\ \ref{figeffaction}, indicating the power of the
2PI formalism.

\begin{figure}[ht]
\centerline{
\epsfysize=1cm\epsfbox{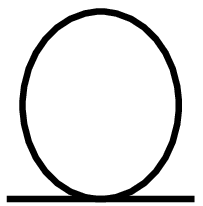}
\hspace*{0.1cm}
\epsfysize=1.2cm\epsfbox{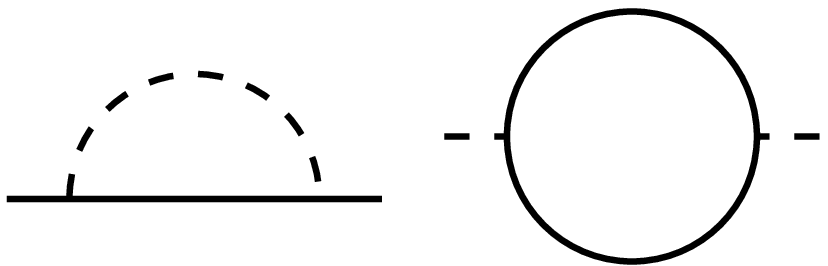}
\hspace*{1.3cm}
\epsfysize=1.2cm\epsfbox{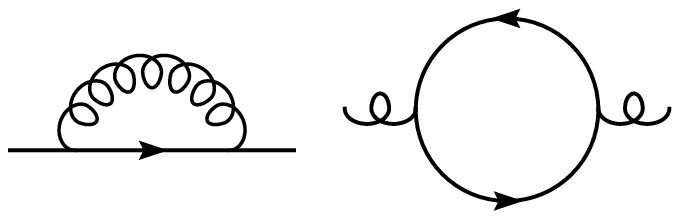}
}   
\caption{Self energies at LO and NLO in the 
large $N$ limit in the $O(N)$ model (left) and at NLO in large $N_f$ QCD 
(right). 
}
\label{figselfenergy}
\end{figure}

Extremizing the effective action yields the Schwinger-Dyson equations for the 
two-point functions. For the scalar case we find
\be
G^{-1} = G_0^{-1} - \Sigma, \;\;\;\;\;\;\;\;D^{-1} = D_0^{-1} -\Pi,
\ee
where $G$ is the scalar and $D$ the auxiliary correlator. For the gauge 
theory
\be
S^{-1} = S_0^{-1} - \Sigma, \;\;\;\;\;\;\;\;D^{-1} = D_0^{-1} -\Pi,
\ee
 with $S$ the fermion and $D$ the gauge field propagator. The self 
energies, depending on
full propagators, are shown in Fig.\ \ref{figselfenergy}.
 In the computation of transport coefficients, it is crucial to use dressed
propagators for two reasons: to screen the so-called pinching
poles,\cite{Lebedev:1990kt,Jeon:if} reflecting the dependence on the
finite lifetime of quasiparticles due to collisions in the plasma, and 
to screen the divergences due to the exchange of offshell gauge 
bosons\cite{baym} (this only in the gauge theory).

Differentiating the self energies yields the rungs that appear in the
Bethe-Salpeter equation. We find that some rungs contribute to the
transport coefficients at leading order, whereas other rungs are
subleading and can be neglected. When considering the Bethe-Salpeter
equation for other purposes (e.g.\ for renormalization\cite{vanHees:2001ik,Blaizot:2003br,Cooper:2004rs}) we
expect that all rungs that follow from the effective action might have to be
considered.

\begin{figure}[ht]
\centerline{
\epsfysize=1.2cm\epsfbox{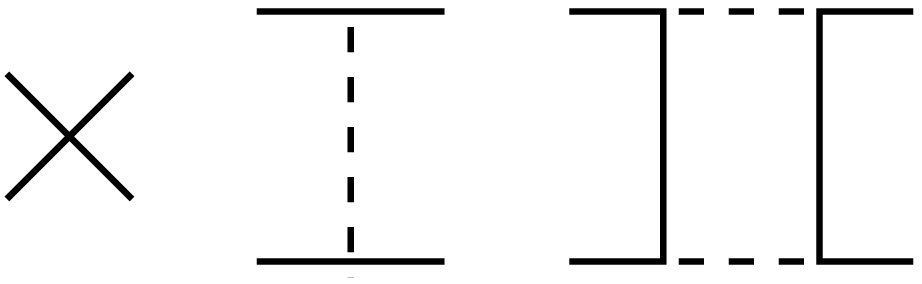}
\hspace*{1.3cm}
\epsfysize=1.2cm\epsfbox{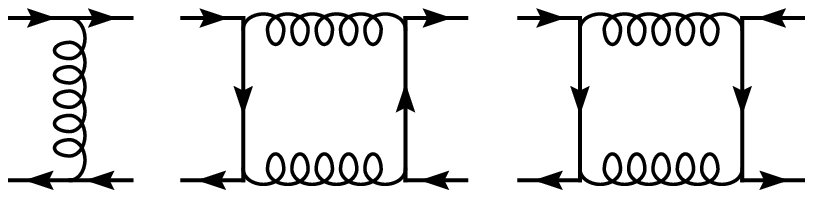}
}   
\caption{Kernels at LO and NLO in the 
large $N$ limit in the $O(N)$ model (left) and at NLO in large $N_f$ QCD 
(right). 
}
\label{figkernel}
\end{figure}

 In order to distinguish leading from subleading rungs, we need to discuss
power counting. In the large $N$ expansion, this is fairly
straightforward. Positive powers of $N$ arise from two sources:  closed
loops of scalar or fermionic lines and scalar or fermionic propagators
suffering from pinching poles (pinching poles are screened by the scalar
or fermionic thermal width $\Gamma_\pv$ and result in contributions $\sim
1/\Gamma_\pv \sim N$). Negative powers of $N$ arise from the coupling
constants, which are taken to scale as $1/N$ as is usually done in $1/N$
expansions. As a result we find the rungs given in Fig.\ \ref{figkernel}.
In the $O(N)$ model, the first point-like rung does not contribute for
kinematical reasons. In large $N_f$ QCD, the contribution from generic
onshell gauge fields is subleading: the gauge field propagator appears
therefore only in rungs and there is no need to consider the thermal width
of onshell gauge bosons. The thermal width of an onshell fermion
determined by the self energy in Fig.\ \ref{figselfenergy} is
ill-defined.\cite{Lebedev:1990kt,Blaizot:1996az} However, the problematic
part cancels against part of the contribution from the line diagram in the
kernel.\cite{Lebedev:1990kt,ValleBasagoiti:2002ir} Finally, we note that
the two box diagrams in large $N_f$ QCD differ only in the orientation of
the fermion lines; this ensures that Furry's theorem is satisfied.


\section{Summation of ladders}

\begin{figure}[ht]
\centerline{
\epsfysize=1.2cm\epsfbox{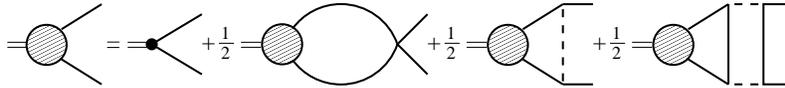}
}
\caption{Integral equation for the effective vertex in  
the $O(N)$ model. The first closed loop on the RHS does not contribute.
}
\label{figintegralON}
\end{figure}

\begin{figure}[ht]
\centerline{
\epsfxsize=4.45in\epsfbox{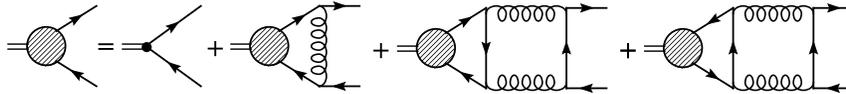}
}   
\caption{Integral equation for the effective vertex in large $N_f$ QCD. In 
the case of the electrical conductivity, the last two contributions 
cancel.
}
\label{figintegralNfQCD}
\end{figure}

To sum the ladder series, we use the technique recently presented by Valle
Basagoiti,\cite{ValleBasagoiti:2002ir} employing the Matsubara formalism
and an effective three-point vertex ${\mathcal D}$. In terms of this
vertex, the integral equations are shown in Figs.\ \ref{figintegralON},
\ref{figintegralNfQCD}.
 Note that the propagators in these diagrams are still dressed. We
consider these equations in the kinematical configuration special for
transport coefficients: the momentum entering on the left is $Q=(q^0\to 0,
\vecnul)$, while the momentum entering and leaving on the right is
$P=(\om_\pv, \pv)$, where $\om_\pv=\sqrt{\pv^2+m^2}$ with $m$ the
scalar or fermion mass.  Transport coefficients are then extracted from
the correlator obtained by closing the lines with an insertion of the
appropriate current, indicated with the small black
dot.\cite{ValleBasagoiti:2002ir,Aarts:2003bk,Aarts:2004sd,Aarts:2004}

The basic quantity we use to solve the integral equations is the ratio of 
the effective vertex ${\mathcal D}(p)$ and the thermal width $\Gamma_\pv$. 
For the shear viscosity we define 
\be
 \chi(p) = \frac{p^2}{\om_\pv} \frac{{\mathcal D}(p)}{\Gamma_\pv} \times 
 \begin{cases}
 1 & O(N) \\
 C_F/T_F & \text{large $N_f$ QCD} 
 \end{cases}
\ee
with $C_F=(N_c^2-1)/2N_c$, $T_F=1/2$. Introducing the color factors in 
this way removes all color factors from the integral equation. 
The shear viscosity in large $N_f$ QCD then follows from a one-dimensional 
integral, 
\be
 \eta = - \frac{d_F T_F}{C_F} \frac{2N_f}{15\pi^2} 
 \int_0^\infty dp\, 
 \frac{p^4}{\om_\pv}n_F'(\om_\pv) \chi(p)
\ee
with $d_F=N_c$. In terms of $\chi$ the integral equations read 
compactly as
\be
 {\mathcal F}(p)\chi(p) = {\mathcal S}(p) +\int_0^\infty dr\, {\mathcal 
 H}(p,r)\chi(r),
\ee
where ${\mathcal F}(p)\propto \Gamma_\pv$, ${\mathcal S}(p)$ is determined 
by the bare vertex, and ${\mathcal H} = {\mathcal H}_{\rm line} + 
{\mathcal H}_{\rm box}$ is determined by the rungs. 
Since ${\mathcal H}(p,r) = {\mathcal H}(r,p)$, the integral equations 
follow as the extremum condition from the functional 
\be
 Q[\chi] = \int_0^{\infty} dp\left[
 {\mathcal S}(p) \chi(p) -\half {\mathcal F}(p)\chi^2(p)+\half
 \int_0^\infty dr\, {\mathcal H}(p,r)\chi(r)\chi(p) \right],
\ee
which allows for a variational treatment. The value of $Q$ at the 
extremum is immediately proportional to the transport coefficient.


\section{Gauge fixing and Ward identity}

In applications of 2PI effective action techniques to gauge theories, 
gauge fixing and Ward identities have to be
considered.\cite{Arrizabalaga:2002hn,Mottola:2003vx,Carrington:2003ut}  
It is therefore interesting to see where gauge fixing parameters appear
and why they drop out in the calculation. We use the generalized Coulomb
gauge such that the gauge field propagator,
\be
D^{\mu\nu} = \Delta_T P^{\mu\nu}_T + \Delta_L g^{\mu 0}g^{\nu 0} + \xi 
\frac{P^\mu P^\nu}{p^4}, 
\ee
consists of a transverse, a longitudinal, and a gauge fixing piece.
The gauge fixing parameter $\xi$ appears in three places:
 \begin{enumerate}
 \item thermal width. The imaginary part of the fermionic self energy is 
proportional to the discontinuity (spectral density) of the 
gauge boson propagator, i.e.\ to $\rho_T$, $\rho_L$. Since the gauge 
fixing part has no discontinuity, $\xi$ drops out.
 \item line diagram. In the kinematical limit we consider also this 
contribution is proportional to the discontinuity of the gauge boson 
propagator and $\xi$ drops out.
 \item box diagram. In the kinematical limit we consider {\em all} fermion 
lines are onshell and $\xi$ drops out exactly.
\end{enumerate}

One can also verify that the effective vertex in Fig.\
\ref{figintegralNfQCD} and the fermion self energy are related via the 
Ward identity, again in the kinematical limit we consider. The
analysis in this case is in fact easier than for the weakly coupled
case,\cite{Aarts:2002tn} since the contribution to the thermal width
arising from soft fermions in the latter is now subleading in the large $N_f$
expansion.\cite{Aarts:2004}

When extending this approach to dynamics far from equilibrium it is
necessary to include contributions that could be dropped in the leading
order analysis of transport coefficients presented here. The gauge fixing
dependence that will be present in that case should be suppressed for
sufficiently large $N_f$.


\section{Variational solution}

The integral equations are in general too complicated to be solved 
analytically. In the scalar case however, we found an approximate but 
surprisingly accurate solution for zero mass and vanishing coupling 
in the limit of ultrahard momentum $p \gg T$. 
We find
\be
\lim_{p\to\infty} \chi(p) = 6912 \pi\frac{N^2}{N+2}\frac{1}{\lambda^2}
\frac{p^2}{T^2}.
\ee
Using this result yields for the shear viscosity in the weakly 
coupled $O(N)$ model,
\be
\label{eqana}
\eta_\infty = \frac{27648\zeta(5)}{\pi}
\frac{N^3}{N+2}\frac{T^3}{\lambda^2}
\approx 3041.9  \frac{3N^3}{N+2}\frac{T^3}{\lambda^2}.
\ee 
 In order to obtain the full $N$ dependence and not just the leading 
order behavior $\sim N^2$ in the large $N$ limit, we used the three-loop 
expansion of the 2PI effective action in the $O(N)$ 
model.\cite{Aarts:2004sd} The result for the numerical prefactor is  
very close to the full results obtained numerically by
Jeon\cite{Jeon:if} (3040) and Arnold, Moore and Yaffe\cite{Arnold:2000dr} 
(3033.5) for $N=1$.

For arbitrary values of the mass and coupling constant (limited by the
presence of the Landau pole), we solve the integral equations
variationally. The function $\chi$ is expanded in a set of trial functions
and the remaining integrals (a one-dimensional integral for ${\mathcal
S}$, three-dimensional integrals for ${\mathcal F}$ and ${\mathcal H}_{\rm
line}$, and a four-dimensional integral for ${\mathcal H}_{\rm box}$) are
performed numerically. We found that a set of four trial functions 
suffices.

\begin{figure}[t]
\centerline{\epsfysize=7cm\epsfbox{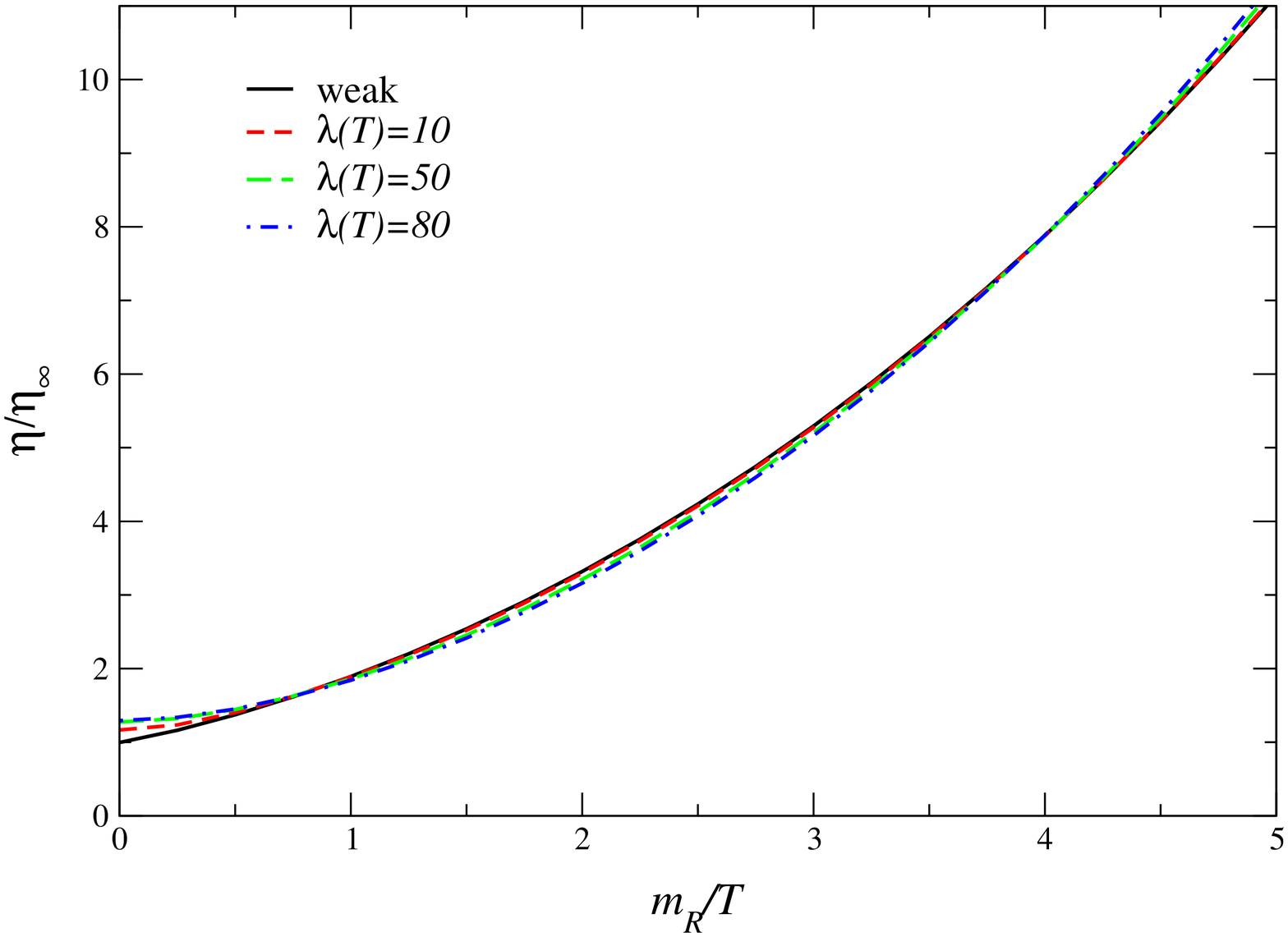}}   
\caption{Shear viscosity in the $O(N)$ model vs.\ the renormalized 
mass at zero temperature for various values of the coupling constant 
$\lambda(\mu=T)$.
}
\label{figshear1}
\end{figure}

The shear viscosity in the $O(N)$ model, normalized with the analytical
result (\ref{eqana}) in the large $N$ limit, is shown in Fig.\
\ref{figshear1} as a function of renormalized mass at zero temperature for
various values of the running coupling constant (the result is
renormalization group independent). We find that the shear viscosity
increases monotonically with increasing mass. This behavior was also found
by Jeon\cite{Jeon:if} for $N=1$. Interestingly, we find surprisingly 
little dependence (apart from that contained in $\eta_{\infty}$) 
on the coupling constant, for all values allowed by the 
Landau pole.

\begin{figure}[t]
\centerline{\epsfysize=7cm\epsfbox{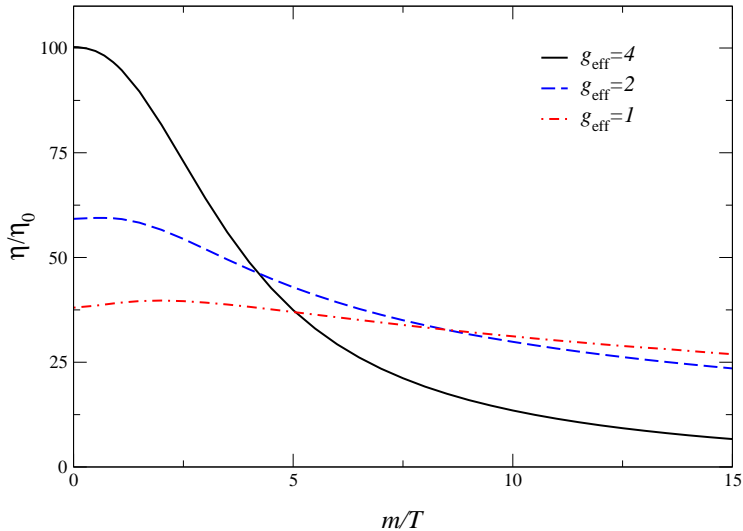}}   
\caption{Shear viscosity in large $N_f$ QCD vs.\ the fermion mass for 
various values of the effective coupling constant $g_{\rm 
eff}(\mu=\pi\exp^{-\gamma_E}T)$.
}
\label{figshear2}
\end{figure}

For large $N_f$ QCD the shear viscosity is shown as a function of fermion
mass in Fig.\ \ref{figshear2}, again for several values of the effective
coupling constant $g_{\rm eff}$ ($g^2_{\rm eff} = T_F g^2 N_f$). The
viscosity is normalized with $\eta_0 = d_F/(T_FC_F)\times T^3/g^4$.  In
this case we observe a nontrivial dependence on the mass. After a slight
increase, we find that the viscosity decreases with increasing mass. This
behavior is caused by longitudinal gauge bosons below the lightcone. In
the limit of very large mass it can be shown\cite{Moore} that the
viscosity goes to zero.\footnote{This is not in contradiction with the
conjectured lower bound on the viscosity/entropy
ratio,\cite{Kovtun:2003wp} since the entropy goes to zero exponentially.}
We also find a much stronger dependence on the coupling constant, compared
to the scalar theory.

\section{Conclusion}

The diagrammatic calculation of transport coefficients in the $O(N)$ model
and large $N_f$ QCD is well organized when the 2PI effective action is
used: all necessary summations are generated by the few diagrams in Fig.\
1. In the scalar theory the shear viscosity increases monotonically 
with mass and has a dominant $1/\lambda^2$ dependence on the coupling
constant. In the gauge theory, on the other hand, the dependence on the
coupling constant and the fermion mass is highly nontrivial, due to 
longitudinal gauge bosons below the lightcone.
 
We emphasize that understanding transport coefficients diagrammatically
provides necessary insight in the dynamics of quantum
fields out of equilibrium. Our results provide further support for the
applicability of truncations of the 2PI effective action to nonequilibrium
QFT.


\section*{Acknowledgments}

We thank Guy Moore for discussions. 
This work was supported by the DOE (Contract No.\ DE-FG02-01ER41190
and DE-FG02-91-ER4069), the Basque Government and in part by the Spanish
Science Ministry (Grant FPA 2002-02037) and the University of the Basque
Country (Grant UPV00172.310-14497/2002).

\end{document}